\documentclass[final]{aipproc}
\layoutstyle{6x9}
\usepackage{epsf}

\begin{document}

\begin{flushright}
CLNS~03/1841
\end{flushright}
\vspace{-0.8cm}

\title{A mixing-independent construction of the unitarity triangle}

\author{Matthias Neubert}{%
address={F.R. Newman Laboratory for Elementary-Particle Physics\\
Cornell University, Ithaca, NY 14853, USA}}

\begin{abstract}
The study of charmless hadronic two-body decays of $B$ mesons is one of 
the most fascinating topics in $B$ physics. A construction of the 
unitarity triangle based on such decays is presented, which is 
independent of $B$--$\bar B$ and $K$--$\bar K$ mixing. It provides 
stringent tests of the Standard Model with small theoretical 
uncertainties.
\end{abstract}

\maketitle

\section{Introduction}

\begin{figure}[h]
\resizebox{8.7cm}{!}{\includegraphics{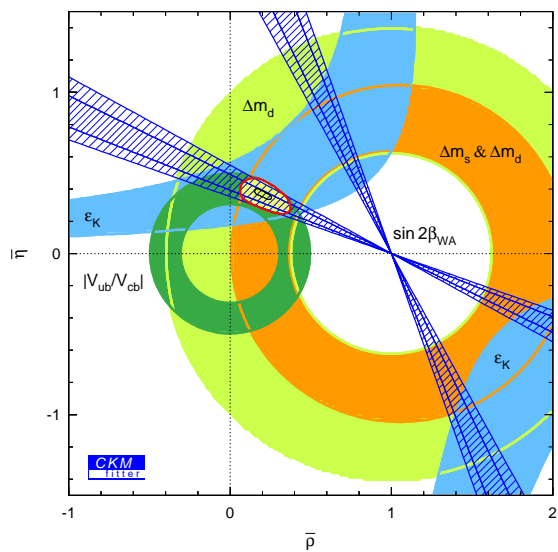}}
\caption{\label{fig:UTfit}
Standard constraints on the apex $(\bar\rho,\bar\eta)$ of the unitarity 
triangle \protect\cite{Hocker:2001xe}.}
\end{figure}

Measurements of $|V_{ub}|$ in semileptonic decays, $|V_{td}|$ in 
$B$--$\bar B$ mixing, and $\mbox{Im}(V_{td}^2)$ from CP violation in 
$K$--$\bar K$ and $B$--$\bar B$ mixing have firmly established the 
existence of a CP-violating phase in the CKM matrix. The present 
situation, often referred to as the ``standard analysis'' of the 
unitarity triangle, is summarized in Figure~\ref{fig:UTfit}. 

Three comments are in order concerning this analysis:
\begin{enumerate}
\item
The measurements of CP asymmetries in kaon physics ($\epsilon_K$ and 
$\epsilon'/\epsilon$) and $B$--$\bar B$ mixing ($\sin2\beta$) probe the 
imaginary part of $V_{td}$ and so establish CP violation in the top 
sector of the CKM matrix. The Standard Model predicts that the imaginary 
part of $V_{td}$ is related, by three-generation unitarity, to the 
imaginary part of $V_{ub}$, and that those two elements are (to an 
excellent approximation) the only sources of CP violation in 
flavor-changing processes. In order to test this prediction one must 
explore the phase $\gamma=\mbox{arg}(V_{ub}^*)$ in the bottom sector of 
the CKM matrix. 
\item
With the exception of the $\sin2\beta$ measurement the standard analysis 
is limited by large theoretical uncertainties, which dominate the widths 
of the various bands in the figure. These uncertainties enter via the 
calculation of hadronic matrix elements. Below I will discuss some novel 
methods to constrain the unitarity triangle using charmless hadronic $B$ 
decays, which are afflicted by smaller hadronic uncertainties and hence 
provide powerful new tests of the Standard Model, which can complement 
the standard analysis.
\item
With the exception of the measurement of $|V_{ub}|$ the standard 
constraints are sensitive to meson--antimeson mixing. Mixing amplitudes 
are of second order in weak interactions and hence might be most 
susceptible to effects from physics beyond the Standard Model. The new 
constraints on $(\bar\rho,\bar\eta)$ discussed below allow a construction 
of the unitarity triangle that is over-constrained and independent of 
$B$--$\bar B$ and $K$--$\bar K$ mixing. It is in this sense complementary 
to the standard analysis.
\end{enumerate}

The phase $\gamma$ can be probed via tree--penguin interference in decays 
such as $B\to\pi K,\pi\pi$. Experiment shows that amplitude interference 
is sizable in these decays. Information about $\gamma$ can be obtained 
from measurements of direct CP asymmetries ($\sim\sin\gamma$), but also 
from the study of CP-averaged branching fractions ($\sim\cos\gamma$). The 
challenge is, of course, to gain theoretical control over the hadronic 
physics entering the tree-to-penguin ratios in the various decays. 
Recently, much progress has been made toward achieving that goal.

Hadronic weak decays simplify greatly in the heavy-quark limit 
$m_b\gg\Lambda_{\rm QCD}$. The underlying physics is that a fast-moving 
light meson produced by a point-like source (the effective weak 
Hamiltonian) decouples from soft QCD interactions 
\cite{Bjorken:1989kk,Politzer:1991au}. A systematic implementation of 
this color transparency argument is provided by the QCD factorization 
approach \cite{Beneke:1999br,Beneke:2001ev}, which makes rigorous 
predictions for hadronic $B$-decay amplitudes in the heavy-quark limit. 
One can hardly overemphasize the importance of controlling nonleptonic 
decay amplitudes in the heavy-quark limit. While a few years ago reliable 
calculations of such amplitudes appeared to be out of reach, we are now 
in a situation where hadronic uncertainties enter only at the level of 
power corrections suppressed by the heavy $b$-quark mass. 

In recent work, QCD factorization has been applied to the entire set of
the 96 decays of $B$ and $B_s$ mesons into $PP$ or $PV$ final states 
($P=$\,pseudoscalar meson, $V=$\,vector meson) \cite{Beneke:2003zv}. It 
has been demonstrated that the approach correctly reproduces the main
features seen in the data, such as the magnitudes of the various tree
and penguin amplitudes, and the fact that they have small relative 
strong-interaction phases. In the future, when more data become available,
this will allow us to extract much useful information about the flavor 
sector of the Standard Model either from global fits or from analysis of 
certain classes of decay modes such as $B\to\pi K$, $B\to\pi\pi$, and 
$B\to\pi\rho$. Detailed comparison with the data may also reveal 
limitations of the heavy-quark expansion in certain modes, perhaps 
hinting at the significance of some power corrections in 
$\Lambda_{\rm QCD}/m_b$.

\section{The CP-b triangle}

Despite of the success of QCD factorization in describing the data, there 
is an interest in analyzing CKM parameters using methods that rely as 
little as possible on an underlying theoretical framework. In this talk I 
discuss a method for constructing the unitarity triangle from $B$ physics 
using measurements whose theoretical interpretation is ``clean'' in the 
sense that it only relies on assumptions that can be tested using 
experimental data. I call this construction the CP-$b$ triangle, because 
it probes the existence of a CP-violating phase in the $b$ sector of the 
CKM matrix. The CP-$b$ triangle is over-determined and can be constructed 
using already existing data. Most importantly, this construction is 
insensitive to potential New Physics effects in $B$--$\bar B$ or 
$K$--$\bar K$ mixing. The present analysis is an update of 
\cite{Neubert:2002tf} using the most recent data as of summer 2003.

\begin{figure}
\resizebox{15cm}{!}{\includegraphics{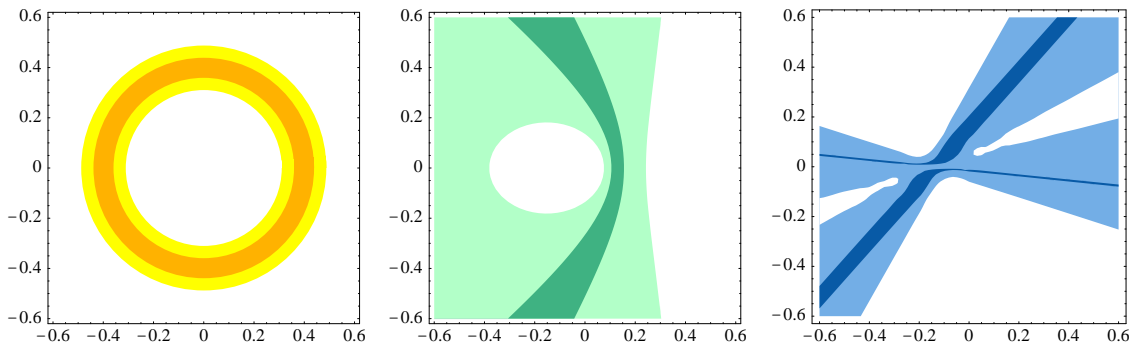}}
\caption{\label{fig:CPT}
The three constraints in the $(\bar\rho,\bar\eta)$ plane used in the 
construction of the CP-$b$ triangle (see text for explanation). 
Experimental errors are shown at 95\% CL. In each plot, the dark band 
shows the theoretical uncertainty, which is much smaller than the 
experimental error. This demonstrates the great potential of these 
methods once the data will become more precise.}
\end{figure}

The first ingredient is the ratio $|V_{ub}/V_{cb}|$ extracted from 
semileptonic $B$ decays, whose current value is 
$|V_{ub}/V_{cb}|=0.09\pm 0.02$. Several strategies have been proposed 
to determine $|V_{ub}|$ with an accuracy of about 10\%
\cite{Neubert:1993um,Dikeman:1997es,Falk:1997gj,Bauer:2000xf,Neubert:2000ch}, 
which would be a significant improvement. The first plot in 
Figure~\ref{fig:CPT} shows the corresponding constraint in the 
$(\bar\rho,\bar\eta)$ plane. Here and below the narrow, dark-colored 
band shows the theoretical uncertainty, while the lighter band gives the 
current experimental value.

The second ingredient is a constraint derived from the ratio of the
CP-averaged branching fractions for the decays $B^\pm\to\pi^\pm K_S$ and 
$B^\pm\to\pi^0 K^\pm$, using a generalization of the method suggested
in \cite{Neubert:1998pt}. The experimental inputs to this analysis are a
certain tree-to-penguin ratio $\varepsilon_{\rm exp}=0.197\pm 0.016$ and 
the ratio
\[
   R_* = 
   \frac{\mbox{Br}(B^+\to\pi^+ K^0)+\mbox{Br}(B^-\to\pi^-\bar K^0)}
        {2[\mbox{Br}(B^+\to\pi^0 K^+)+\mbox{Br}(B^-\to\pi^0 K^-)]}
   = 0.804\pm 0.085
\]
of two CP-averaged $B\to\pi K$ branching fractions \cite{newdata}. 
Without any recourse to QCD factorization this method provides a bound 
on $\cos\gamma$, which can be turned into a determination of 
$\cos\gamma$ (for fixed value of $|V_{ub}|/V_{cb}|$) when information 
on the relevant strong-interaction phase $\phi$ is available. The phase 
$\phi$ is bound by experimental data (and very general theoretical 
arguments) to be small, of order $10^\circ$ \cite{Neubert:2002tf}. (In 
the future, this phase can be determined directly from the direct CP 
asymmetry in $B^\pm\to\pi^0 K^\pm$ decays.) It is thus conservative to 
assume that $\cos\phi>0.8$, corresponding to $|\phi|<37^\circ$. With this 
assumption the corresponding allowed region in the $(\bar\rho,\bar\eta)$ 
plane was analyzed in \cite{Beneke:2001ev}. The resulting constraint is 
shown in the second plot in Figure~\ref{fig:CPT}. 

The third constraint comes from a measurement of the time-dependent
CP asymmetry $S_{\pi\pi}=-\sin2\alpha_{\rm eff}$ in $B\to\pi^+\pi^-$ 
decays. The present experimental situation is still unclear, since the 
measurements by BaBar ($S_{\pi\pi}=-0.40\pm 0.22\pm 0.03$) and Belle 
($S_{\pi\pi}=-1.23\pm 0.41_{\,-0.07}^{\,+0.08}$) are not in good 
agreement with each other \cite{LP03}. The naive average of these 
results gives $S_{\pi\pi}=-0.58\pm 0.20$. (Inflating the error according 
to the PDG prescription would yield $S_{\pi\pi}=-0.58\pm 0.34$, but for 
some reason the experimenters usually use the naive error without 
rescaling, and I will follow their example.) The theoretical expression 
for the asymmetry is
\[
   S_{\pi\pi} = - \frac{2\,\mbox{Im}\,\lambda_{\pi\pi}}
                       {1+|\lambda_{\pi\pi}|^2} \,,
   \quad \mbox{where} \quad
   \lambda_{\pi\pi} = e^{-i\phi_d}\,
    \frac{e^{-i\gamma} + (P/T)_{\pi\pi}}
         {e^{+i\gamma} + (P/T)_{\pi\pi}} \,.
\]
Here $\phi_d$ is the CP-violating phase of the $B_d$--$\bar B_d$ mixing
amplitude, which in the Standard Model equals $2\beta$. Usually it is
argued that for small $(P/T)_{\pi\pi}$ ratio the quantity 
$\lambda_{\pi\pi}$ is approximately given by 
$e^{-2i(\beta+\gamma)}=e^{2i\alpha}$, and so apart from a ``penguin 
pollution'' the asymmetry $S_{\pi\pi}\approx-\sin2\alpha$. In order to 
become insensitive to possible New Physics contributions to the mixing 
amplitude I adopt a different strategy \cite{Beneke:2001ev}. I use the 
measurement $\sin\phi_d=0.736\pm 0.049$ \cite{LP03} and write 
$e^{-i\phi_d}=\pm(1-\sin^2\!\phi_d)^{1/2}-i\sin\phi_d$, with a sign 
ambiguity in the real part. (The plus sign is suggested by the standard
fit of the unitarity triangle.) A measurement of $S_{\pi\pi}$ can then 
be translated into a constraint on $\gamma$ (or $\bar\rho$ and 
$\bar\eta$), which remains valid even if the $\sin\phi_d$ measurement 
is affected by New Physics. The result obtained with the current 
experimental values and assuming $\cos\phi_d>0$ is shown in the third 
plot in Figure~\ref{fig:CPT}. The resulting bands for $\cos\phi_d<0$ 
are obtained by a reflection about the $\bar\rho$ axis. This follows
because the expression for $S_{\pi\pi}$ is invariant under the 
simultaneous replacements $e^{-i\phi_d}\to -e^{i\phi_d}$ and 
$\gamma\to -\gamma$.

\begin{figure}
\resizebox{14.2cm}{!}{\includegraphics{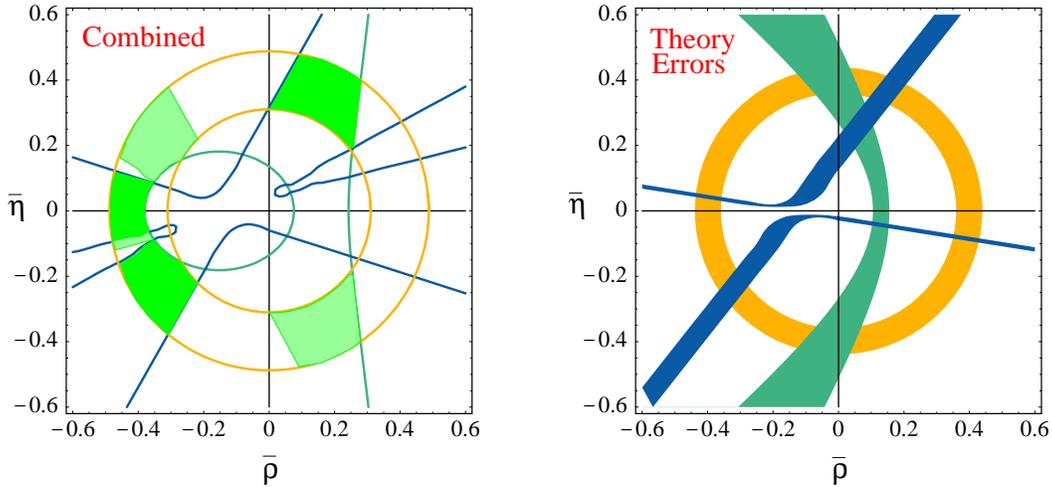}}
\caption{\label{fig:summary}
Left: Allowed regions in the $(\bar\rho,\bar\eta)$ plane obtained from 
the construction of the CP-$b$ triangle (at 95\% CL). The light-shaded 
areas refer to $\cos\phi_d<0$. Right: Theoretical error bands for the 
three constraints combined in the construction of the CP-$b$ triangle.}
\end{figure}

Each of the three constraints in Figure~\ref{fig:CPT} are, at present, 
limited by rather large experimental errors, while comparison with 
Figure~\ref{fig:UTfit} shows that the theoretical limitations are 
smaller than for the standard analysis. Yet, even at the present level
of accuracy it is interesting to combine the three constraints and 
construct the resulting allowed regions for the apex of the unitarity 
triangle. The result is shown in the left-hand plot in 
Figure~\ref{fig:summary}. Note that the lines corresponding to the new 
constraints intersect the circles representing the $|V_{ub}|$ constraint 
at large angles, indicating that the three measurements used in the 
construction of the CP-$b$ triangle provide highly complementary 
information on $\bar\rho$ and $\bar\eta$. There are six (partially 
overlapping) allowed regions, three corresponding to $\cos\phi_d>0$ (dark 
shading) and three to $\cos\phi_d<0$ (light shading). If we use the 
information that the measured value of $\epsilon_K$ requires a positive 
value of $\bar\eta$, then only the solutions in the upper half-plane 
remain. Comparison with Figure~\ref{fig:UTfit} shows that one of these 
regions (corresponding to $\cos\phi_d>0$) is in perfect agreement with 
the standard fit. This is highly nontrivial, since with the exception of 
$|V_{ub}|$ none of the standard constraints are used in this 
construction. Interestingly, there is a second allowed region 
(corresponding to $\cos\phi_d<0$) which would be consistent with the 
constraint from $\epsilon_K$ but inconsistent with the constraints 
derived from $\sin2\beta$ and $\Delta m_s/\Delta m_d$. Such a solution 
would require a significant New Physics contribution to $B$--$\bar B$ 
mixing.

\vspace{0.2cm}
{\em Acknowledgment:\/}
This research was supported by the National Science Foundation under 
Grant PHY-0098631.

\bibliographystyle{aipproc}

\end{document}